\documentclass[
    ,final            
  ]
  {aipproc}

\layoutstyle{6x9}

\usepackage{psfig}

\def\lya{\ifmmode {\rm Ly}\alpha~ \else Ly$\alpha$~\fi}
\def\lyan{\ifmmode {\rm Ly}\alpha \else Ly$\alpha$\fi}
\def\lyb{\ifmmode {\rm Ly}\beta~ \else Ly$\beta$~\fi}
\def\lyg{\ifmmode {\rm Ly}\gamma~ \else Ly$\gamma$~\fi}
\def\civ{\ifmmode {\rm C}\,{\sc iv}~ \else C\,{\sc iv}~\fi}
\def\civn{\ifmmode {\rm C}\,{\sc iv}~ \else C\,{\sc iv}\fi}
\def\cvi{\ifmmode {\rm C}\,{\sc vi}~ \else C\,{\sc vi}~\fi}
\def\cvin{\ifmmode {\rm C}\,{\sc vi} \else C\,{\sc vi}\fi}
\def\hi{H\,{\sc i}~}
\def\mgii{\ifmmode {\rm Mg}\,{\sc ii}~ \else Mg\,{\sc ii}~\fi}
\def\mgiin{\ifmmode {\rm Mg}\,{\sc ii} \else Mg\,{\sc ii}\fi}
\def\cvin{\ifmmode {\rm Mg}\,{\sc ii} \else Mg\,{\sc ii}\fi}
\def\niv{\ifmmode {\rm N}\,{\sc iv}~ \else N\,{\sc iv}~\fi}
\def\nivn{\ifmmode {\rm N}\,{\sc iv} \else N\,{\sc iv}\fi}
\def\nv{N\,{\sc v}~}
\def\nvn{N\,{\sc v}}

\def\ovi{{{\rm O}\,{\sc vi}~}}

\def\chandra {{\it Chandra}~}

\def\chandran {{\it Chandra}}

\begin{document}

\title{Metallicity measurements in AGNs}

\classification{90}
\keywords {galaxies:abundances--galaxies:Seyfert--quasars:absorption lines--quasars:emission lines--quasars:individual (Mrk 1044, Mrk 279)--ultraviolet:galaxies}

\author{Smita Mathur}{
  address={The Ohio State University}
}

\author{Dale Fields}{
  address={Pierce College}
}


\begin{abstract}
Measuring metallicity in the nuclear regions of AGNs is difficult
because only a few lines are observed and ionization correction
becomes a major problem. Nitrogen to carbon ratio has been widely used
as an indicator for metallicity, but precise measurements have been
lacking. We made such measurements for the first time using a wide
baseline of ionization states with observations from FUSE, HST and
\chandran. \ovi observations with FUSE were crucial in this effort. We
measured super-solar metallicities in two AGNs and found that N/C does
not scale with metallicity. This suggests that chemical enrichment
scenario in nuclear regions of galaxies may be different from
traditional models of metal enrichment.
\end{abstract}

\maketitle


\section{Metallicity in AGNs}

Understanding chemical evolution of galaxies is one of the fundamental
problems in astronomy and active galaxies are no exception. We now
understand that most, if not all galaxies exhibit an active phase some
times in their life. Thus knowing metallicity in AGNs allows us
understand the chemical enrichment in galactic nuclei which may differ
substantially from the enrichment on galactic scales. AGNs also show
signs of outflow. The role of such outflows, especially those from
powerful quasars, in enriching the intergalactic medium with metals is
unclear. How does the kinematics, dynamics and metal content of AGN
outflows compare with those from super-winds on galactic scales? While
these are all important questions to answer, we have no knowledge of
metallicity in the circumnuclear regions of AGNs.

This is because measuring AGN metallicity is difficult. In the
optical, where most of the measurements are made, only a handful of
broad emission lines are observed, mostly of hydrogen. At high
redshift, the rest-frame UV spectrum is observed with several metal
lines such as \civn, \nvn, \mgiin. However, converting the emission
line strengths to metallicity is not straight forward as it depends
upon ionization balance, temperature, density and
geometry. Nonetheless, attempts to estimate metallicities have been
made, notably by Hamann \& Ferland (1999 and references therein). They
suggested the use of \nvn/\civ as a metallicity indicator. Because of
the secondary C-N-O nucleosynthesis, N/C scales as metallicity Z, so N
scales like Z$^2$. Using these method, Hamann
\& Ferland showed that high redshift quasars have super-solar
metallicity.

The use of emission lines as metallicity indicators was questioned by
Shemmer \& Netzer (2002) who showed that \nivn/\civ ratio and
\nvn/\civ ratio do not give consistent answers for N/C! Given the
strong model dependence of emission line strengths, this was not a
surprise. Absorption line strengths, on the other hand, are geometry
and density independent and so are potentially better tracers of
metallicity. Converting the observed column densities of absorption
lines to column densities of metals still involves ionization
correction; once again this is a difficult quantity to measure because
only a handful of lines of different metals are observed in a given
band. A long base-line of ionization states is required to make the
ionization correction. As discussed below, we performed
multi-wavelength observations toward this goal and made the first
precise measurement of metallicity in an AGN.

\section{Metallicity measurement in Mrk 1044}


\begin{figure}[ht]
\psfig{file=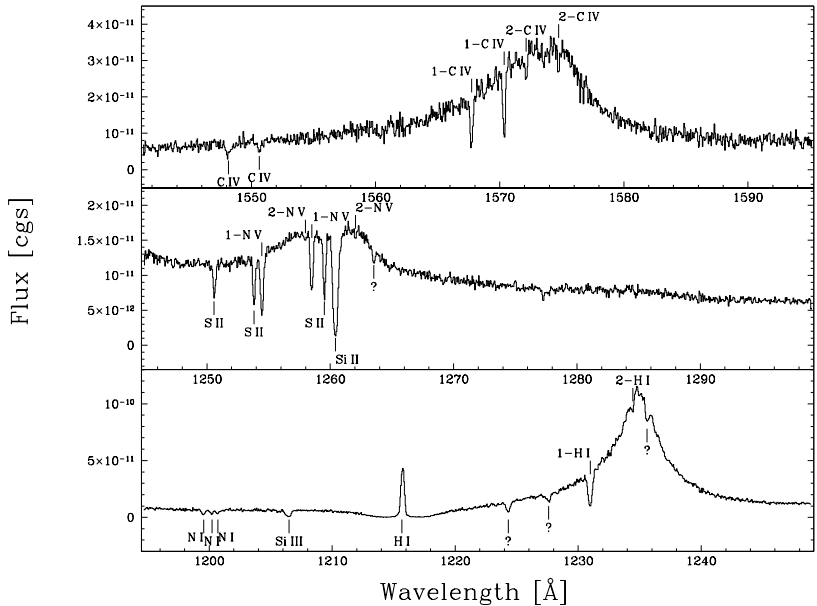,height=3.5in,angle=-90}
\caption{HST STIS spectrum of Mrk 1044. Note the absorption lines of
 \civn, \nvn, and \lyan.}
\end{figure}

Figure 1 shows the HST STIS spectrum of Mrk 1044. Multiple velocity
components of \civ, \nv and \lya are clearly seen.  A FUSE spectrum of
Mrk 1044 is shown in figure 2. Here we see absorption lines of \ovi at
the same velocities seen in the HST spectrum. Additionally we detect
Ly$\beta$ absorption. We fit a ``pseudo continuum'' through the
emission lines and the AGN continuum and measure the strengths of the
absorption lines and determine their column densities (see Fields et
al. 2005a,b for details).

The next step is to determine the ionization parameter of the absorption
system. To this end we generated a grid of models of ionization
parameter $\rm U$ and the total column density $N_H$ for solar
metallicity and mixture using CLOUDY and looked for models consistent
with observations. This is shown in figure 3; the hatched curves
correspond to the locus on the $\log N_H--\log \rm U$ plane corresponding to
the observed column density of each ion. The intersection of all the
curves, if present, defines the $N_H$ and U of the system. The curves
of \civn, \nv and \ovi actually do intersect at $\log \rm U =-1.29$. What is
noteworthy, however, is that the curves for \hi do not. For the
inferred value of $\rm U$, the \hi column density is significantly
lower. This implies that the metals are more abundant, or that the
metallicity of the system is super-solar. We measure the metallicity
to be about 5 times solar. 

What is also interesting is that the metal mixture is consistent with
solar. In particular, we do not find the locus of \nv displaced from
the intersection of \civ and \ovi. Thus, we find no evidence for N
scaling as $Z^2$.

\begin{figure}[t]
\psfig{file=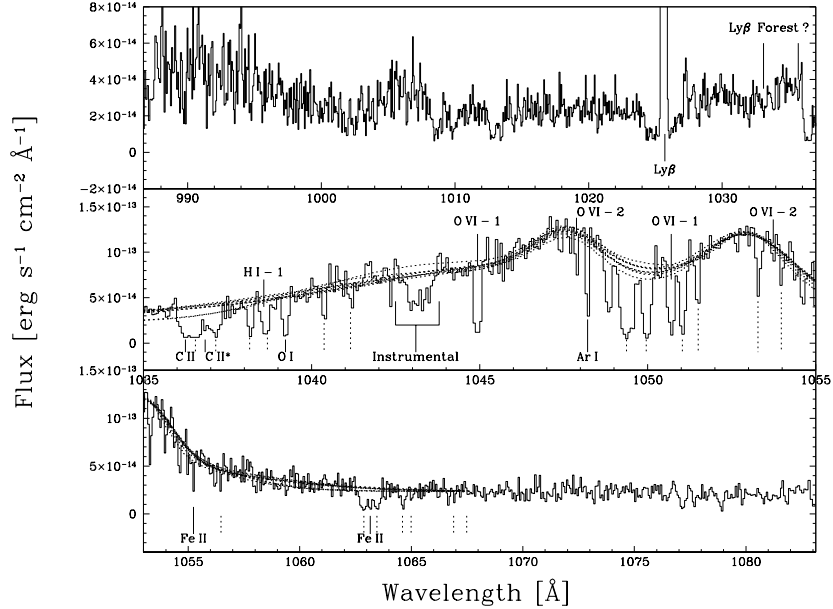,height=3.5in,angle=-90}
\caption{FUSE spectrum of Mrk 1044. Note the absorption lines of \ovi
 and Ly$\beta$ marked above the continuum fit including emission
 lines.}
\end{figure}

\section{Metallicity in Mrk 279}

Encouraged by the success in measuring metallicity in Mrk 1044, we
attempted to apply our technique to another AGN for which
multi-wavelength observations existed. Mrk 279 was observed with
\chandran, HST and FUSE. In contrast to UV/FUV bands, the X-ray region
contains hundreds of lines spanning a range of ionization states and
so potentially provides powerful diagnostics. We found that the
\chandra spectrum of Mrk 279 did not have sufficiently high quality to
make robust metallicity measurements. However, the data were better
fit with a model containing super-solar metallicity. 

A consistent picture is now emerging with high resolution grating
spectroscopy of X-ray absorbing outflows. The absorber seems to be
made up of two or more components in pressure balance with each
other. Costantini et al. (2007) had reported that the two components
in Mrk 279, the one with low ionization parameter (LIP) and the one
with high ionization parameter (HIP) are not in pressure
equilibrium. In fact the pressure--temperature curve in Costantini et
al. did not have an equilibrium zone. We had shown previously that
super-solar metallicity can restore an equilibrium zone in the
pressure--temperature curve (Komossa \& Mathur 2000). So we re-made a
pressure--temperature plot for the absorbers in Mrk 279 with
super-solar metallicity as suggested by the fit to the \chandra
data. As shown in figure 5, this not only restored a equilibrium zone,
but now the LIP and the HIP components were found to be in pressure
balance. While not conclusive, this make the case for super-solar
metallicity in Mrk 279 stronger. 

\begin{figure}[t]
\psfig{file=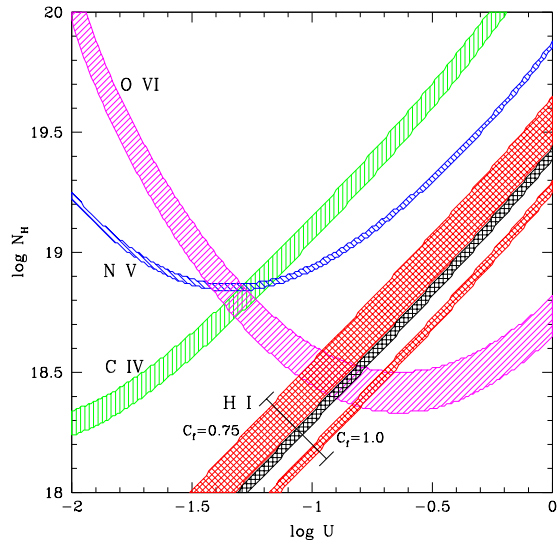,height=3in}
\caption{CLOUDY models assuming solar metallicity. Shaded regions
 correspond to observed column densities at the 1$\sigma$ level for
 several ions. Note that all the metal lines agree in a small region
 of parameter space around $\log U= -1.29$ and $\log N_H= 18.85$. The
 distance in $\log N_H$ between this point and the HI curve is at least
 $+0.7$.}
\end{figure}

We then compared our X-ray results with the UV data from Gabel et
al. (2005). Only when we invoke the model with super-solar
metallicity, could we match the X-ray and UV data. Given that the LIP
X-ray component is generally found to be consistent with the UV
absorber, this gives an additional supporting evidence for super-solar
metallicity in Mrk 297 (Fields et al. 2007). Arav et al (2007) independently
arrived at the same conclusion.

\section{Conclusions}

Studies of Mrk 1044 and Mrk 279 have given strong evidence for
super-solar metallicity in these AGNs. The Mrk 1044 result with five
times solar metallicity is robust; first such precise measurement for
an AGN. While we find metallicity to be overall super-solar, the
abundance mixture is consistent with solar. Notably, we do not find
any evidence for N/C scaling with metallicity. These studies imply
that the chemical enrichment process in galactic nuclei is likely to
be very different from traditional models. The Cosmic Origins
Spectrograph (COS) to be installed on HST will prove to be invaluable
in generalizing these results for a large number of AGNs. 

\begin{figure}[t]
\psfig{file=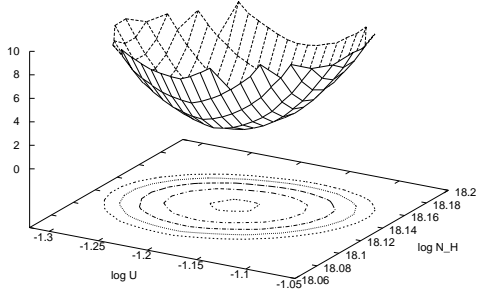,width=3in,angle=-90}
\caption{The $\chi^2$ surface for models for bulk metallicity of 5
 times solar, with a solar mixture of metals and helium enhanced by
 $\Delta Y/\Delta z= 2$ (see Fields et al. 2005b for details).}
\end{figure}

\begin{figure}[b]
\psfig{file=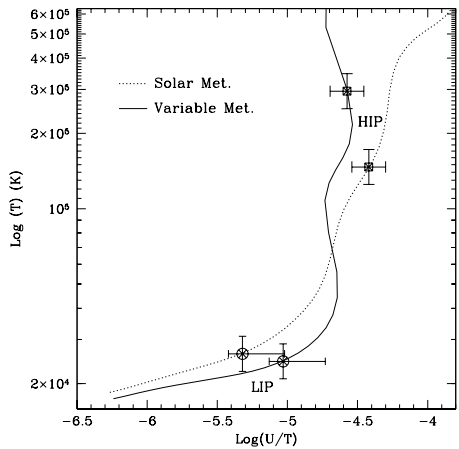,height=2.5in}
\caption{The pressure--temperature plot for Mrk 279. The dotted curve
 is for solar metallicity while the solid curve is for super-solar
 C,N,O as observed. The LIP and the HIP components do not correspond
 to the same pressure (plotted as $\log U/T$) on the dotted curve, but
 are consistent with the same pressure on the solid curve. Note also
 that there is no equilibrium zone in the dotted curve, but there is
 one in the solid curve.}
\end{figure}


  



\bibliographystyle{aipprocl} 


\begin{thebibliography}{}

\bibitem{Brown2000}
N. Arav, et al. 
\emph{ApJ}, \textbf{658}, 829 (2007).

\bibitem{BrownAustin:2000}
E. Costantini, et al. \emph{A\&A} \textbf{461}, 121
   (2007).

\bibitem{Wang}
D. Fields, S. Mathur, R. Pogge, F. Nicastro, and St. Komossa, \emph{ApJ}, \textbf{620}, 183 (2005a)

\bibitem{SJ:1999}
D. Fields, S. Mathur, R. Pogge, F. Nicastro, St. Komossa, and Y. Krongold \emph{ApJ}, \textbf{634}, 928  (2005b)

\bibitem{}
D. Fields, S. Mathur, Y. Krongold, Rik Williams, and  F. Nicastro,  \emph{ApJ}, \textbf{666}, 828  (2007)

\bibitem{}
F. Hamann \& G. Ferland, \emph{ARA\&A}, \textbf{37}, 487 (1999)

\bibitem{}
St. Komossa and  S. Mathur, \emph{A\&A}, \textbf{374}, 914 (2001)

\bibitem{}
O. Shemmer, and H. Netzer, \emph{ApJ}, \textbf{567}, L19 (2002)

\end{thebibliography}




\end{document}